\documentclass[aps, prb, showpacs, 10pt, twocolumn, superscriptaddress, floatfix, longbibliography, nofootinbib]{revtex4-1}

\usepackage{graphicx,amsfonts,amssymb,amsmath,xspace,dsfont}
\usepackage[colorlinks=true,citecolor=green,linkcolor=red,urlcolor=blue]{hyperref}
\usepackage{bbm}

\newcommand{\bra}[1]{\langle#1|}
\newcommand{\ket}[1]{|#1\rangle}
\newcommand{\id}{\mathbbm{1}}

\begin{document}

\title{Tensor-network approach to phase transitions in string-net models}

\author{Alexis Schotte} 
\email{alexis.schotte@ugent.be}
\affiliation{Department of Physics and Astronomy, Ghent University, Krijgslaan 281, 9000 Gent, Belgium}

\author{Jose Carrasco}
\affiliation{Departamento de F\'\i sica Te\'orica, Universidad Complutense de Madrid, 28040 Madrid, Spain} \affiliation{Institute for Theoretical Physics, University of Innsbruck, Technikerstrasse 21A, 6020 Innsbruck, Austria}

\author{Bram Vanhecke}
\affiliation{Department of Physics and Astronomy, Ghent University, Krijgslaan 281, 9000 Gent, Belgium}

\author{Laurens Vanderstraeten}
\affiliation{Department of Physics and Astronomy, Ghent University, Krijgslaan 281, 9000 Gent, Belgium}

\author{Jutho Haegeman}
\affiliation{Department of Physics and Astronomy, Ghent University, Krijgslaan 281, 9000 Gent, Belgium}

\author{Frank Verstraete}
\affiliation{Department of Physics and Astronomy, Ghent University, Krijgslaan 281, 9000 Gent, Belgium}

\author{Julien Vidal} 
\email{vidal@lptmc.jussieu.fr}
\affiliation{Sorbonne Universit\'e, CNRS, Laboratoire de Physique Th\'eorique de la Mati\`ere Condens\'ee, LPTMC, F-75005 Paris, France}

\begin{abstract}
 We use a recently proposed class of tensor-network states to study phase transitions in string-net models. These states encode the genuine features of the string-net condensate such as, e.g., a nontrivial perimeter law for Wilson loops expectation values, and a natural order parameter detecting the breakdown of the topological phase. In the presence of a string tension, a quantum phase transition occurs between the topological phase and a trivial phase. We benchmark our approach for $\mathbb{Z}_2$ string nets and capture the second-order  phase transition which is well known from the exact mapping onto the transverse-field Ising model. More interestingly, for Fibonacci string nets, we obtain first-order transitions in contrast with previous studies but in qualitative agreement with mean-field results.
\end{abstract}

\maketitle

\section{Introduction} 

Since its discovery in the late 1980s, topological order aroused much interest in physics. The long-range entanglement structure as well as the exotic quasiparticle excitations associated with this order may prove essential in attempts to achieve scalable fault-tolerant quantum computers or quantum memories~\cite{Nayak08}. As such, it is of paramount importance to understand how perturbations generate dynamics and interactions between the anyonic excitations and induce a breakdown of topological phases.

\par One of the most famous models hosting topological order was proposed by Levin and Wen in 2005~\cite{Levin05}. The string-net Hamiltonian allows to describe all doubled achiral topological phases. Thus, it has been the starting point for many studies concerning phase transitions~\cite{Gils09_1,Gils09_3,Ardonne11,Burnell11_2,Schulz13,Morampudi14,Schulz14,Schulz15,Schulz16_1,Schulz16_2,Burnell18,Vidal18}. Nevertheless, in the absence of local order parameter, the nature of these transitions remains an open question in many cases since one cannot use Landau's theory of symmetry breaking.  

\par Among all alternative methods developed to study these topological phase transitions, a particularly versatile framework for constructing variational states is provided by tensor networks. In two dimensions, the projected entangled-pair states (PEPSs)~\cite{Verstraete04} are known to describe the string-net ground states~\cite{Gu09,Chen10_2} and directly encode the topological properties in the virtual symmetries of the local PEPS tensor~\cite{Schuch10,Sahinoglu14,Bultinck2017}. This feature has been exploited to detect possible topological phase transitions and to identify the associated anyon-condensation mechanism~\cite{Burnell18} in Abelian~\cite{Haegeman15_1,Haegeman15_2,Shukla18,Zhu2019} and non-Abelian~\cite{Marien17,Xu2019} cases at the level of wave functions. However, variational PEPS calculations for concrete models have so far been restricted to $\mathbb{Z}_N$ toric  codes~\cite{Gu08,Dusuel11,Schulz12} whose excitations are Abelian anyons. 
Recently, a family of PEPS based on perturbative expansions has been introduced to describe different ground states across a phase transition~\cite{Vanderstraeten17}. 

For a given Hamiltonian that exhibits a phase transition, the procedure to build these ``perturbative PEPSs"  can be summarized as follows: (i) we start from a wave function that describes the phase transition at the mean-field level; (ii) we apply tensor-network operators which implement the perturbative expansions in an extensive way to wave functions on both sides of the transition; (iii) we promote the {\it ad hoc} coefficients of these expansions to variational parameters. In two dimensions, these perturbatively exact variational states are still PEPSs and the tensor-network machinery~\cite{Fishman18, Vanderstraeten16} can be used to perform an efficient variational optimization. In Ref.~\onlinecite{Vanderstraeten17}, this method has been notably applied to the $\mathbb{Z}_2$ toric code perturbed with a string tension~\cite{Trebst07,Hamma08} for which the virtual symmetry of the local PEPS tensor emerges as an order parameter.

\par In this work, we go one step beyond and implement a variational   PEPS to study phase transitions in both Abelian ($\mathbb{Z}_2$) and non-Abelian (Fibonacci) string-net models on the the honeycomb lattice. For the $\mathbb{Z}_2$ case, we capture the second-order quantum phase transition known from the mapping onto the transverse-field Ising model on the triangular lattice~\cite{Burnell11_2}. In the Fibonacci case, we only find first-order phase transitions, in contrast to Ref.~\onlinecite{Schulz13} but in agreement with the mean-field results~\cite{Dusuel15}.

\section{String-net models} We consider the two-dimensional string-net model introduced by Levin and Wen~\cite{Levin05} in the presence of a tension term. For simplicity, we focus on the simplest case where the microscopic degrees of freedom, defined on the links of a honeycomb lattice, can only be in two different states, 0 and 1 (when possible, we omit the ket notation to describe states). The Hilbert space $\mathcal{H}$ is defined as the set of configurations obeying the branching rules that stem from the fusion rules of the theory considered~\cite{Levin05}. 

In the present work, we discuss two different theories, $\mathbb{Z}_2$ and Fibonacci, whose fusion rules are given by:
%
%
\begin{eqnarray}
\mathbb{Z}_2\quad &:&   0 \times a= a \times 0=a,  1\times1=0,  \label{eq:Z2}\\
\rm{Fibonacci}&:&  0 \times a= a \times 0=a, 1\times1=0+1,  \label{eq:Fib}
\label{eq:fusion}
\end{eqnarray}
%
%
for $a=0,1$. As underlined in Ref.~\onlinecite{Levin05}, there are actually two different theories obeying $\mathbb{Z}_2$ fusion rules that give rise to either a doubled $\mathbb{Z}_2$ (D$\mathbb{Z}_2$)  or a doubled  semion (Dsem) topological phase. For the string tension considered thereafter, phase diagrams are the same for both theories.

At each vertex of the honeycomb lattice, the fusion rules must be satisfied~\cite{Levin05}, i.e., if two links are in states $a$ and $b$ the third link must be in a state $c \in a\times b$. Following Ref.~\onlinecite{Simon13}, one can compute the dimension of the Hilbert space. For any trivalent graph with $N_\mathrm{v}$ vertices, one then gets
%
%
\begin{eqnarray}
\mathbb{Z}_2\quad &:&  \dim \mathcal{H}= 2^{\frac{N_\mathrm{v}}{2}+1}, \quad\\
\rm{Fibonacci}&:&  \dim \mathcal{H}= (1+\varphi^2)^{\frac{N_\mathrm{v}}{2}}+(1+\varphi^{-2})^{\frac{N_\mathrm{v}}{2}}, \quad
\label{eq:dimH}
\end{eqnarray}
%
%
where $\varphi=\frac{1+\sqrt{5}}{2}$ is the golden ratio. 

In order to analyze the breakdown of the topological phase originating from the string-net model, we consider the following Hamiltonian:
%
\begin{equation}
H= - J_\mathrm{p}\sum_p B_p -J_\mathrm{l}\sum_l L_l.
\label{eq:ham}
\end{equation}
%
%
The first term corresponds to the usual string-net Hamiltonian introduced by Levin and Wen in Ref.~\onlinecite{Levin05}. Operators $B_p$'s are mutually commuting projectors that ``measure'' the flux in the plaquette $p$. The action of $B_p$ on a given link configuration depends on the theory under consideration through its $F$-symbols [see Eq. (C1) in Ref.~\onlinecite{Levin05} for details]. The operator $B_p$ only modifies the six inner links of the plaquette $p$ but its action depends (diagonally) on the six outer links~\cite{Levin05}. For $J_{\mathrm{p}}>0$ and $J_{\mathrm{l}}=0$, all ground states are flux free and hence characterized by $B_p=1$ for all $p$, up to a
topology-dependent degeneracy. 

The second term is also a sum of mutually commuting projectors. Operators  $L_l$'s are diagonal in the canonical (link) basis and act as $L_l {\ket a}_l= \delta_{a,0} {\ket a}_l$, where ${\ket a}_l$ denotes the state of the link $l$.
For $J_\mathrm{l}>0$, this second term favors configurations with links in the state $0$ and penalizes strings of links in the state $1$, hence the name string tension.

\par For $J_\mathrm{l}>0$ and $J_\mathrm{p}=0$, the ground state is unique (trivial phase) and given by the product state \mbox{${\ket 0}= \otimes_l {\ket 0}_l$} for {\em both} $\mathbb{Z}_2$ and Fibonacci fusion rules.  For $J_\mathrm{l}<0$ and $J_\mathrm{p}=0$, the ground-state manifold depends on the fusion rules. Indeed, for Fibonacci fusion rules, the product state  ${\ket 1}= \otimes_l {\ket 1}_l$ is allowed and is  the unique ground state. By contrast, for $\mathbb{Z}_2$ fusion rules, this state is not allowed (since $1\times 1=0$) and the ground space is spanned by all allowed states with $N_{\mathrm{v}}$ links in the state $1$ and $\frac12 N_{\mathrm{v}}$ links in the state $0$.

\section{Methodology} 

The goal of this work is to analyze phase transitions from the topological phase existing for $J_\mathrm{p}>0$ in the small $|J_\mathrm{l}/J_\mathrm{p}|$ limit to the trivial phases found in the large $|J_\mathrm{l}/J_\mathrm{p}|$ limit. To this aim, let us set $J_\mathrm{p}=\cos\theta$ and ${J_\mathrm{l}=\sin\theta}$ and consider first the region where $\theta \in [0,\pi/2]$. Following the variational tensor-network approach introduced in Ref.~\onlinecite{Vanderstraeten17}, we consider the state
\begin{equation}
  \ket{\alpha,\beta}=\mathcal{N}  \exp{\Big(\beta \sum_l L_l \Big)}\prod_p(\id+\alpha Z_p)\ket{0},
  \label{eq:state}
\end{equation}
where $Z_p=2 B_p-\id$, $\alpha$ and $\beta$ are variational parameters, and $\mathcal{N}$ is a normalization factor. 
According to Ref.~\onlinecite{Vanderstraeten17}, a better description of the trivial phase would be obtained by adding an extra term $\exp (-\gamma \sum_p B_p)$. However, it considerably increases the complexity of the PEPS so that we do not consider it in the following.

For $\beta=0$, the state $\ket{\alpha,0}$ describes the phase transition at the mean-field level~\cite{Dusuel15} and the states $\ket{1,0}$ and $\ket{0,0}$ are the exact ground states for $\theta=0$ and $\theta=\pi/2$, respectively. Furthermore, the first-order contribution in $\alpha$ to $\ket{\alpha,\beta}$ around $(\alpha,\beta)=(0,0)$ corresponds to the first-order perturbative correction to the exact ground state around $\theta=\pi/2$. Likewise, the first-order contribution in $\beta$ to $\ket{\alpha,\beta}$ around $(\alpha,\beta)=(1,0)$ corresponds to the first-order perturbative correction to the exact ground state around $\theta=0$~\cite{Vanderstraeten17}. 

The state $\ket{\alpha,\beta}$ can be interpreted as a PEPS, whose bond dimension depends on the theory considered. The variational energy per plaquette
\begin{equation} \label{eq:e0}
  e_0(\alpha,\beta)=\frac{1}{N_{\rm p}}\frac{\bra{\alpha,\beta} H \ket{\alpha,\beta}}{\langle{\alpha,\beta}\ket{\alpha,\beta}},
\end{equation}
can be efficiently computed using the VUMPS algorithm \cite{Fishman18} for contracting two-dimensional tensor networks in the thermodynamic limit. 
Note that the previous approach reproduces the linear perturbative corrections up to second order both near $\theta=0$ and $\theta=\pi/2$. As explained in Ref.~\onlinecite{Vanderstraeten17}, additional tensor-network operators can be added in order to reproduce higher-order perturbative corrections.

\par The PEPS framework allows for a natural characterization of the topological nature of the variational ground state. Indeed, the topological properties of a PEPS are related to the virtual symmetries of the local PEPS tensor~\cite{Schuch10}. In the Fibonacci theory, this virtual symmetry is described by a matrix product operator (MPO)~\cite{Sahinoglu14, Bultinck2017}. Since, as shown in the Appendix, the state $\ket{\alpha,\beta}$ exhibits such a virtual MPO symmetry only when $\alpha=1$, this parameter can be naturally interpreted as an order parameter~\cite{Vanderstraeten17} to detect the transition between the topological phase ($\alpha=1$) and the trivial one ($\alpha<1$). Indeed, at $\alpha=1$, the expectation value of a Wilson loop operator changes from a trivial perimeter law for $\beta=0$ to a non-trivial perimeter law for $\beta>0$, still indicating deconfinement of the anyonic excitations, so that the state remains in the topological phase. For $\alpha<1$, the Wilson loop expectation value satisfies a non-trivial area law, indicating that anyons are confined and the state is in the trivial phase.

Yet, even in the presence of the virtual symmetry ($\alpha=1$), the parameter $\beta$ can drive the state into a trivial phase by a spontaneous breaking of this symmetry, resulting in an area law for the Wilson loop. This process was shown to occur at $\beta=\frac{1}{2}\ln(1+\sqrt{2})$ for the $\mathbb{Z}_2$ case~\cite{Castelnovo08}) and at $\beta \simeq 0.168$ for the Fibonacci case~\cite{Marien17}. For the problem at hand, we checked that there is no spontaneous symmetry breaking so that $\alpha$ can be used as a bona-fide order parameter.

\section{Results for the \texorpdfstring{$\mathbb{Z}_2$}{Z2} theory}
Let us first discuss the simplest theory and consider $\mathbb{Z}_2$ (Abelian) fusion rules. As discussed in Ref.~\onlinecite{Burnell11_2}, the Hamiltonian (\ref{eq:ham}) for any theory obeying $\mathbb{Z}_N$ fusion rules can be exactly mapped onto the $N$-states Potts model in a magnetic field on the dual lattice. Thus, for $N=2$, $H$ is equivalent to the transverse-field Ising model on a triangular lattice where $J_\mathrm{p}$ and $J_\mathrm{l}$ are the strength of the magnetic field and of the spin-spin coupling, respectively. As a result, the sign of $J_\mathrm{p}$ is irrelevant for this problem and we assume $J_\mathrm{p}>0$ in the following. 

In the antiferromagnetic case ($J_\mathrm{l}<0$), the Ising model on a triangular lattice is highly frustrated. So, clearly, the ansatz $\ket{\alpha,\beta}$ is not adapted to that situation since the ground space has an extensive degeneracy for \mbox{$J_\mathrm{p}=0$}. In this region $\theta \in[-\pi/2,0]$, a critical point in the universality class of the three-dimensional classical $XY$ model was found for \mbox{$\theta \simeq \arctan(-1/65) \simeq -0.545$~\cite{Isakov03}}.

\par
\begin{figure}[t]
  \includegraphics[width=\columnwidth]{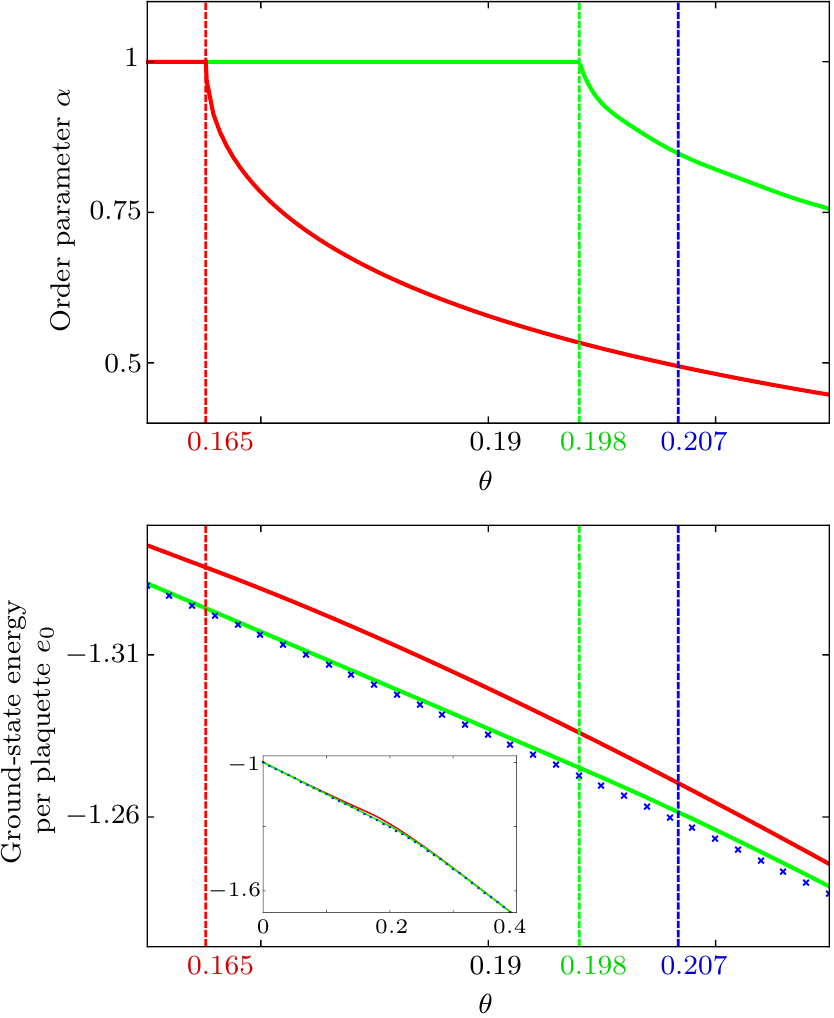}
  \caption{Variational results for the $\mathbb Z_2$ theory obtained
    for \mbox{$\beta=0$} (red) and $\beta \neq 0$ (green).  Top: order
    parameter $\alpha$ indicating a continuous transition from a
    topological phase ($\alpha=1$) to a trivial phase
    ($\alpha<1$). Bottom: ground-state energy per plaquette $e_0$
    compared with exact diagonalization data (blue crosses) (see inset for a broader
    range). Dashed lines give the position of the transition point
    obtained from series expansions~\cite{He90} (blue) and from the
    order parameter behavior [red~\cite{Dusuel15} and green (this work)].}
  \label{fig:Z2}
\end{figure}

 Here, we rather aim at benchmarking our ansatz with the phase transition in the region $\theta \in[0,\pi/2]$ corresponding to ferromagnetic interactions ($J_\mathrm{l}>0$). The phase diagram in this region has been studied by high-order series expansion and a second-order transition occurs at $\theta_{\rm c}\simeq\arctan(0.2097)\simeq0.207$~\cite{He90}, the critical point belonging to the universality class of the three-dimensional classical Ising model. Our results obtained from the variational ansatz (\ref{eq:state}) are summarized in Fig.~\ref{fig:Z2} (top panel). As already discussed in Ref.~\onlinecite{Dusuel15}, for $\beta=0$ (mean-field approximation), one obtains a continuous transition at $\theta=\arctan(1/6)\simeq0.165$ which is qualitatively correct but about $20\%$ off from $\theta_{\rm c}$. Remarkably, by adding $\beta$ as a second variational parameter, the transition remains continuous (no jump of the order parameter $\alpha$) and shifts to $\theta\simeq0.198$ which is only $4\%$ off from $\theta_{\rm c}$. Given that our variational ansatz has only short-ranged correlations [except for $\alpha=1$ and $\beta=\frac{1}{2}\ln(1+\sqrt{2})$, which is not a variational optimum for any value of $\theta$], and that the exact correlations decay algebraically near $\theta_{\mathrm c}$, these results can be considered as unexpectedly good. This can also be seen by comparing the variational ground-state energy with the numerical results obtained from exact diagonalization on a $25$-plaquettes system with periodic boundary conditions, as shown in
Fig.~\ref{fig:Z2} (bottom panel).

Note, finally, that for the $\mathbb{Z}_2$ theory, our ansatz satisfies $\ket{\alpha,\beta} \sim \ket{\alpha^{-1},\beta}$, such that the expansion of the energy density $e_0(\alpha,\beta)$ around $\alpha = 1 + \delta \alpha$ can only contain even terms in $\delta \alpha$. As a consequence, the way $\alpha$ deviates from $1$ around the transition point is as $|\delta \alpha| = (\theta-\theta_c)^{1/2}$, both for the mean-field ansatz (with fixed $\beta = 0$) and for the ansatz where $\beta$ is also optimized. 
%
%
%
%

\section{Results for the Fibonacci theory} 

\begin{figure}[t]
  \includegraphics[width=\columnwidth]{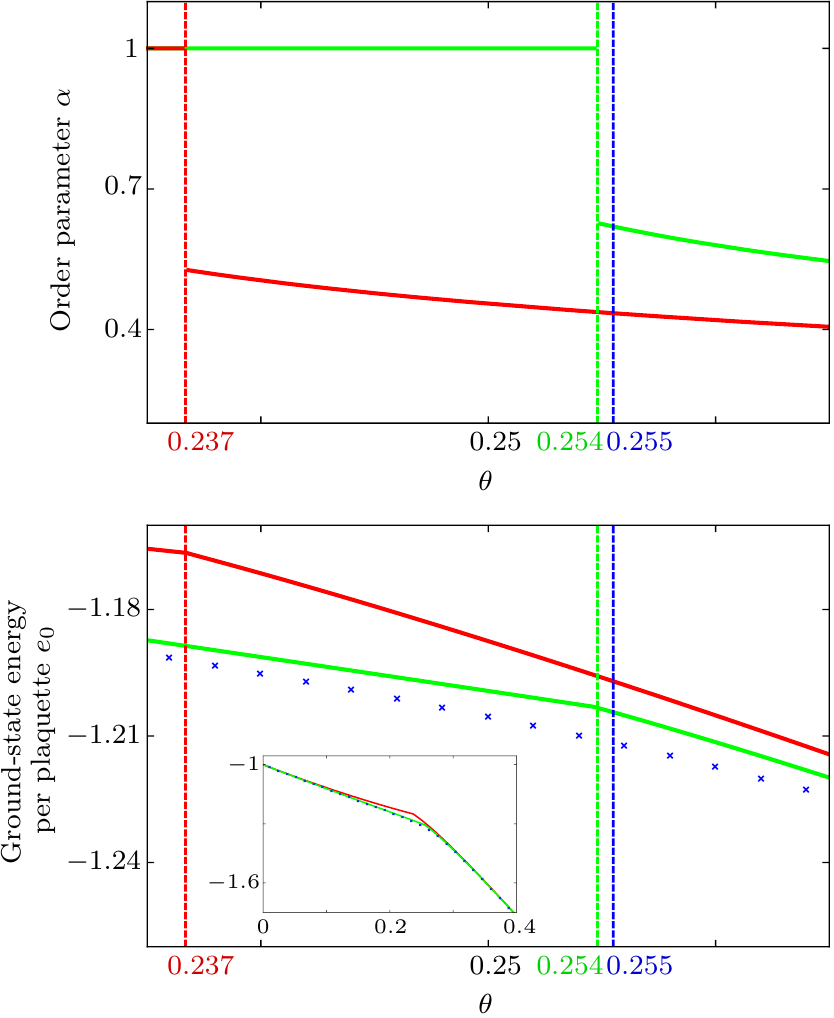}
  \caption{Variational results for the Fibonacci theory (same
    conventions as in Fig.~\ref{fig:Z2}). Blue dashed lines indicate
    the position of the transition point computed from series \mbox{expansions~\cite{Schulz13}}.}
  \label{fig:Fibo_pos}
\end{figure}

\begin{figure}[t]
  \includegraphics[width=\columnwidth]{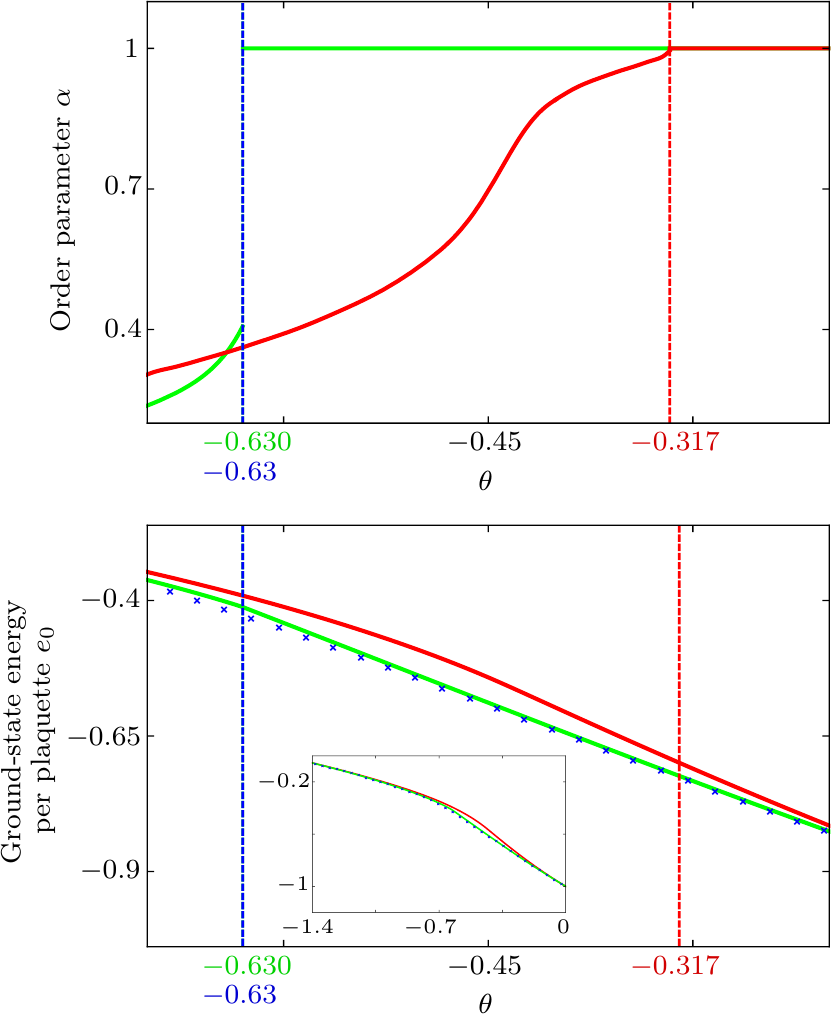}
  \caption{Variational results for the Fibonacci theory (same
    conventions as in Fig.~\ref{fig:Z2}). Blue dashed lines indicate
    the position of the transition point computed from series \mbox{expansions~\cite{Schulz13}}
    which coincides with the one obtained with the ansatz
    (\ref{eq:state2}), i.e., $\theta\simeq -0.630$.}
  \label{fig:Fibo_neg}
\end{figure}

The phase diagram of the Hamiltonian (\ref{eq:ham}) for the Fibonacci theory has already been computed by combining exact diagonalization results with high-order series expansions for the ground-state and gap energies~\cite{Schulz13}. The doubled Fibonacci (DFib) topological phase has been found to extend from~$\theta_2\simeq-0.63$ to~$\theta_1\simeq0.255$ identifying $\theta_1$ and $\theta_2$ as critical points. Our variational results in this case are displayed in Figs.~\ref{fig:Fibo_pos} and~\ref{fig:Fibo_neg} for $\theta\in[0,\pi/2]$ and $\theta\in[0,-\pi/2]$, respectively. 

\par In the region $\theta\in[0,\pi/2]$, the mean-field approach~\cite{Dusuel15} corresponding to $\beta=0$ indicates a first-order transition for~$\theta=\arctan\left(\frac{1+\varphi}{6+3\varphi}\right)\simeq0.237$. This result is qualitatively different from the one proposed in Ref.~\onlinecite{Schulz13} although the position of the transition point is only $7\%$ off from $\theta_1$. Since (i) there is no prior reason to believe that the mean-field result is exact~\cite{Dusuel15} and (ii) higher-order series expansions need to be extrapolated to provide reliable  informations about the nature of the transition, it is interesting to see what the ansatz (\ref{eq:state}) can bring to the understanding of the transition.  As can be seen in Fig.~\ref{fig:Fibo_pos}, by adding $\beta$ as variational parameter, one still obtains a first-order transition characterized by a jump of the order parameter, but the transition point is shifted to $\theta\simeq 0.254$, which is less than $1\%$ off from $\theta_1$. This leads us to conclude that in the region \mbox{$\theta\in[0,\pi/2]$}, there is a unique transition point located near $\theta\simeq 0.255$ (in agreement with Ref.~\onlinecite{Schulz13}) corresponding to a first-order transition with a small gap at the transition point (weakly first-order). Note that the proximity of the transition points obtained by the two approaches could suggest that the flux-flux correlation length is finite, which is a favorable case for a PEPS description of the ground state and justifies the relevance of our ansatz.
%
%
%
%

In the region $\theta\in[-\pi/2,0]$, we investigate the phase transition by slightly modifying the ansatz. Indeed, the state defined in Eq.~\eqref{eq:state} is  designed for interpolating between $\ket{1,0}$ and $\ket{0,0}$, that are the exact ground states at $\theta=0$ and $\theta=\pi/2$, respectively. However, for $\theta=-\pi/2$, the ground state of the Hamiltonian~\eqref{eq:ham} is unique (topologically trivial phase) and given by~${\ket 1}=\otimes_l{\ket 1}_l$. Consequently, to study the parameter range $\theta\in[-\pi/2,0]$, we consider as variational ansatz
\begin{equation}
  \ket{\alpha,\beta}_{-}=\mathcal{N}  \exp{\Big(\beta \sum_l L_l \Big)}\prod_p(\id+\alpha Z_p)\ket{1},
  \label{eq:state2}
\end{equation}
where, for simplicity, we kept the same notations as in Eq.~(\ref{eq:state}). The PEPS tensor encoding this state is described in the Appendix. It is important to note that the ``reference" states $\ket{0}$ and $\ket{1}$ are very different. Indeed, for~$\beta=0$, a key property of the ansatz~(\ref{eq:state}) is the factorization property that reads
\begin{equation}
  \langle\alpha,0 | \prod_{p=1}^n  B_p |\alpha,0\rangle=\langle \alpha,0 | B_p |\alpha,0\rangle^n, 
\end{equation}
for any set of $n$ plaquettes. This identity does not hold for the state~\eqref{eq:state2} with $\beta=0$, which can no longer be interpreted as a mean-field ansatz. Yet, by construction, it is perturbatively exact near $\theta=0$ and it also matches the exact ground state at $\theta=-\pi/2$. As such, this ansatz is a good candidate to capture the transition between the DFib phase and the trivial phase. As can be seen in Fig.~\ref{fig:Fibo_neg}, for $\beta=0$, one obtains a continuous transition at $\theta\simeq-0.317$ which is very far from the value~$\theta_2\simeq-0.63$ obtained in Ref.~\onlinecite{Schulz13}. Interestingly, when including $\beta$, we obtain a discontinuous transition located at $\theta\simeq-0.630$ which is in agreement with the extrapolated values of Ref.~\onlinecite{Schulz13}. Thus, we face a situation similar to the previous case, where the present variational study is in quantitative agreement with the series expansions studies. We emphasize that the series expansions in this region have to be resummed so that error bars on $\theta_2$ are larger than for $\theta_1$. Regarding the nature of the transition, the same arguments as before favor the first-order scenario.

\section{Conclusions and outlook}  

This work presents the first variational results based on tensor networks for a topological phase transition out of a non-Abelian topological phase. We have studied the transitions between the topological and trivial phases of the Levin-Wen Hamiltonian with string tension, both for $\mathbb{Z}_2$ and Fibonacci fusion rules, by means of a simple two-parameters variational ansatz inspired by Ref.~\onlinecite{Vanderstraeten17}. For the $\mathbb{Z}_2$ case, we recover the well-known second-order transition as predicted from the mapping onto the transverse-field Ising model the triangular lattice. For the Fibonacci case, our results are in quantitative agreement with series expansions and exact diagonalizations~\cite{Schulz13}. However, we only find first-order transitions (as in the mean-field treatment~\cite{Dusuel15}) whereas series expansions combined with exact diagonalizations rather plead in favor of second-order transitions~\cite{Schulz13}. This qualitative discrepancy is likely due to the extrapolation of the series expansion and finite-size effects in the exact diagonalizations, but we cannot exclude that the present variational approach is not sufficient to properly describe the transitions in this model. Going beyond would require more sophisticated ans\"atze that can be systematically constructed by using the ideas developed in Ref.~\onlinecite{Vanderstraeten17}. Using recently developed contraction methods for three-dimensional tensor networks \cite{Vanderstraeten18}, we stress that such an approach can also be applied in three-dimensional systems, as recently illustrated in Ref.~\onlinecite{Reiss19}.

\acknowledgments

JC would like to thank the Laboratoire de Physique Th\'eorique de la Mati\`ere Condens\'ee of the Sorbonne Universit\'e and the Department of Physics and Astronomy of the Ghent University for their warm hospitality. This research was supported by the Research Foundation Flanders, and ERC Grants No. QUTE (647905), No. ERQUAF (715861) and QTFLAG.

\onecolumngrid
\appendix

\section{Construction of the PEPS tensor}
	
In the following section, we elaborate on the construction of the PEPS tensors representing the ansatz \mbox{$\ket{\alpha, \beta}_{+} = \ket{\alpha, \beta}$} and $ \ket{\alpha, \beta}_{-} $.
It is sufficient to construct the PEPS for the one-parameter ans\"atze $ \ket{\alpha, 0}_{\pm}$.
The PEPS tensor for  $ \ket{\alpha, \beta}_{\pm} \propto \prod_l \exp \left (\beta L_l\right )  \ket{\alpha,0}_{\pm}$ is then readily found by applying the right operator on the physical level.
		
For simplicity, we will assume that  quantum dimensions $d_0$ and $d_1$ are non-negative real numbers, and that the \mbox{$F$-symbols} are all real. Both models studied in this work satisfy these assumptions (see Ref.~\onlinecite{Levin05} for details about $\mathbb{Z}_2$ and Fibonacci theories). In the general case, where one relaxes these assumptions, the construction of the PEPS can be done in a similar fashion.
			
	In order to simplify the calculations in the following section, we will work with the states
	\begin{align}\label{key}
		\ket{\gamma}_{\pm} = \ket{ \gamma D^2/(-2\gamma + 2 +\gamma D^2), 0}_{\pm}.
	\end{align}
	To go back to the ansatz used in the main body of the paper, one can simply use
	\begin{equation}\label{key2}
		\ket{\alpha,0}_{\pm} = \ket{2 \alpha/(2\alpha - \alpha D^2+D^2)}_{\pm},
	\end{equation}
	where $D=\sqrt{d_0^2+d_1^2}$ is total quantum dimension.

	\subsection{Ansatz for \texorpdfstring{$\theta \in [0,\pi/2]$}{region 1}}
		The one-parameter ansatz is given by 
		\begin{equation}\label{eq:ansatz}
		\ket{\gamma} = \mathcal{N} \prod_{p} (\id +  \gamma d_1 O_1^p) \ket{0} ,
		\end{equation}
		where $ \ket{0}= \otimes_l \ket{0}_l $, and  $ \mathcal{N} $ is a normalization factor.
		The operator $  O_i^p $ corresponds to inserting a $ i $-loop inside the plaquette \cite{Levin05}, and then resolving it into the lattice using $F$-moves
		\begin{equation} \label{eq:F-move}
			\raisebox{-0.4cm}{\includegraphics[scale=.4]{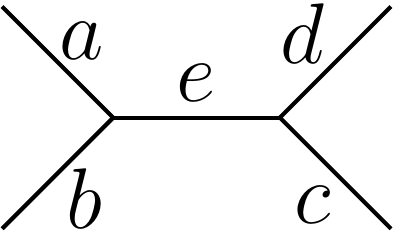}}
			= \;
			\sum_{f} \; F_{cd \, f}^{ab \, e} \;\;
			\raisebox{-.8 cm}{\includegraphics[scale=.4]{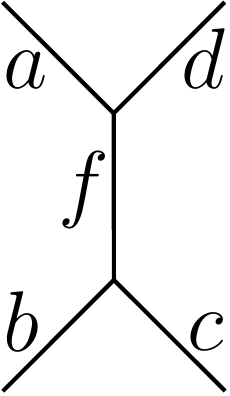}} ,
		\end{equation}
		and the rule
		\begin{equation} \label{eq:quant_dim}
			\raisebox{-0.3cm}{\includegraphics[scale=.38]{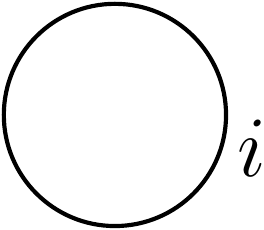}} \;
			= d_i .
		\end{equation}
		Contraction of a loop can not happen across a plaquette, we treat plaquettes as if they have a puncture in their center.
		Applying these rules gives the full form of $ O_i^p $:
		\begin{equation} \label{eq:O}
		\quad O_i^p \, \Bigg |
		\raisebox{-0.6cm}{\includegraphics[scale=.28]{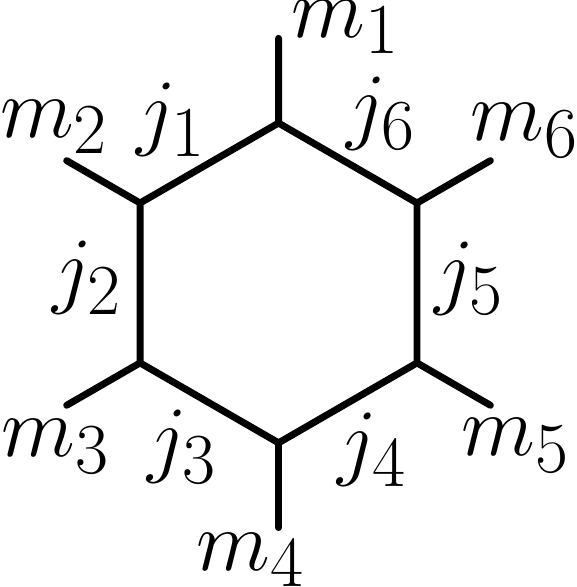}}
		\Bigg \rangle
		= \!\!
		\sum_{k_1,\ldots,k_6} 
		\!\!\!\! \bigg( \! \prod_{\nu = 1}^6 F^{m_\nu j_\nu j_{\nu-1}}_{i k_{\nu-1} k_\nu} \! \bigg)
		\Bigg |
		\raisebox{-0.6cm}{\includegraphics[scale=.28]{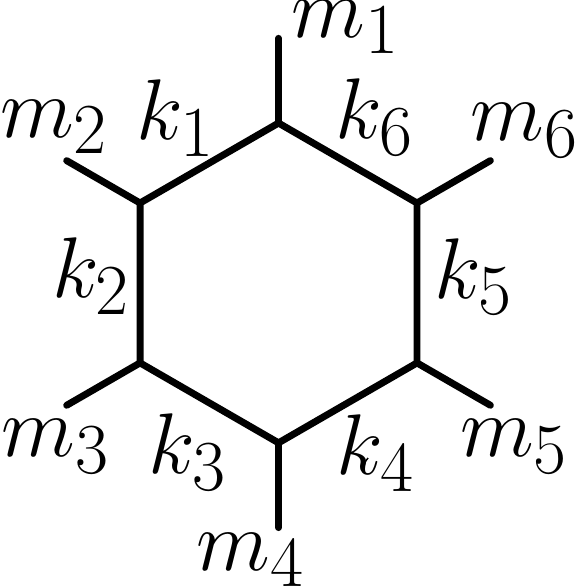}}
		\Bigg \rangle .
		\end{equation}
		Setting $ \tilde{d}_0 = 1$ and $ \tilde{d}_1 = \gamma \, d_1 $ one can write
		\begin{equation}\label{eq:ansatz2}
			\ket{\gamma} = \mathcal{N} \prod_{p} \left ( \sum_{\mu_p} \tilde{d}_{\mu_p} O_{\mu_p}^p \right ) \ket{0} .
		\end{equation}
		Using the graphical representation of the operators  $  O_i^p $, $ \ket{\gamma} $ can be represented as
		\begin{equation}\label{eq:plaquettes_loops}
			\ket{\gamma} =  \mathcal{N} \sum_{\mu_1 , \mu_2 , \dots} \tilde{d}_{\mu_1} \tilde{d}_{\mu_2} \dots \raisebox{-2cm}{\includegraphics[scale=.3]{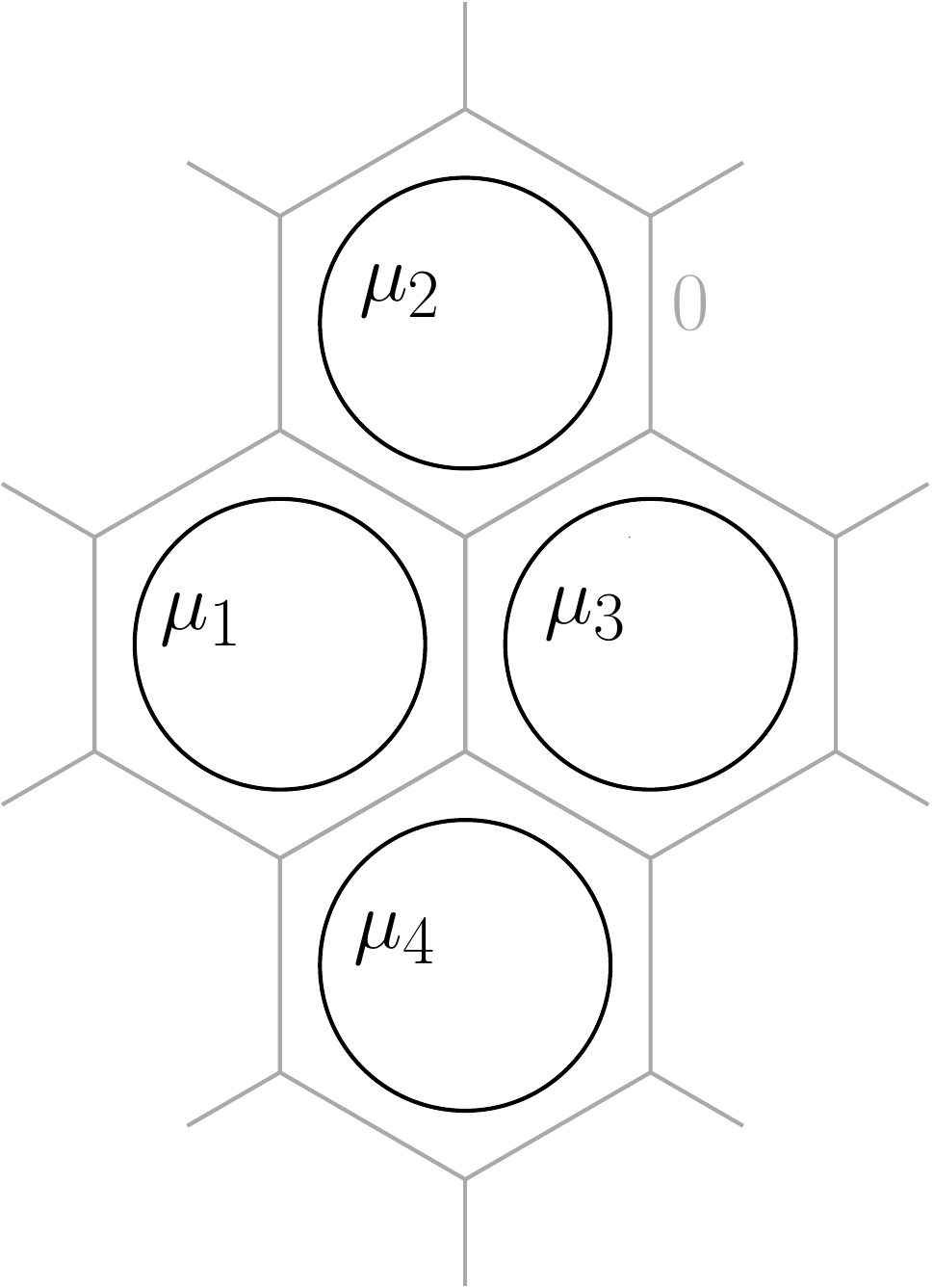}} .
		\end{equation}
		The gray lines above are initially in the $ \ket{0} $ state.
		To find the state from this graphical notation, one has to resolve the loops appearing in Eq.~\eqref{eq:plaquettes_loops} into the lattice.
		This is done in two steps.
		First we fuse the neighboring loops along each edge, using an F-move:
		\begin{equation}\label{eq:munu}
			 \raisebox{-0.4cm}{\includegraphics[scale=.5]{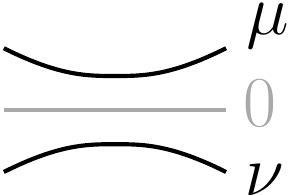}}  = \sum_{k} F^{\mu \mu \, 0}_{\nu\nu \, k} \; \raisebox{-0.4cm}{\includegraphics[scale=.5]{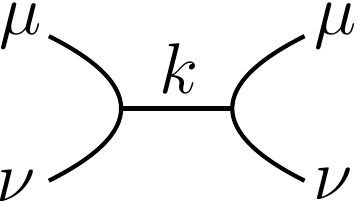}} = \sum_{k} \sqrt{\dfrac{d_k}{d_\mu d_\nu}} \delta_{\mu\nu k} \; \raisebox{-0.4cm}{\includegraphics[scale=.5]{munu2.pdf}} .
		\end{equation}
		The $ k $ appearing in the sum, will be the physical degree of freedom in every edge, once we are done resolving everything into the lattice.
		The second step consists of using the following equality in every vertex:
		\begin{equation}\label{eq:vertex_loop}
			\raisebox{-0.62cm}{\includegraphics[scale=.38]{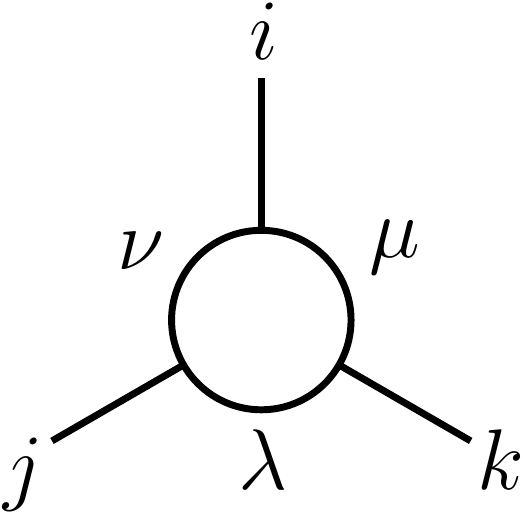}} = \sqrt{d_\lambda d_\mu d_\nu} G^{i j k}_{\lambda \mu \nu} \raisebox{-0.42cm}{\includegraphics[scale=.38]{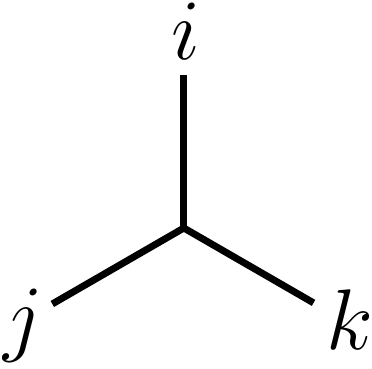}} ,
		\end{equation}
		where 
		\begin{equation}\label{eq:G}
			G^{ijk}_{\lambda \mu \nu} = \dfrac{1}{\sqrt{d_i d_\lambda}} F^{\nu j \lambda}_{k \mu i} = \dfrac{1}{\sqrt{d_\nu d_k}} \left(F^{ijk}_{\lambda \mu \nu}\right)^* .
		\end{equation}
		The PEPS tensor is obtained by splitting the factor $ \sqrt{d_k / d_\mu d_\nu}$ in Eq.~\eqref{eq:munu} evenly between the two adjacent vertices, while splitting the factor $ \tilde{d}_{\mu_p}$ in Eq.~\eqref{eq:ansatz2} evenly between all 6 vertices appearing in plaquette $ p $.
		The result is
		\begin{equation}\label{eq:PEPS}
			 \raisebox{-1cm}{\includegraphics[scale=.42]{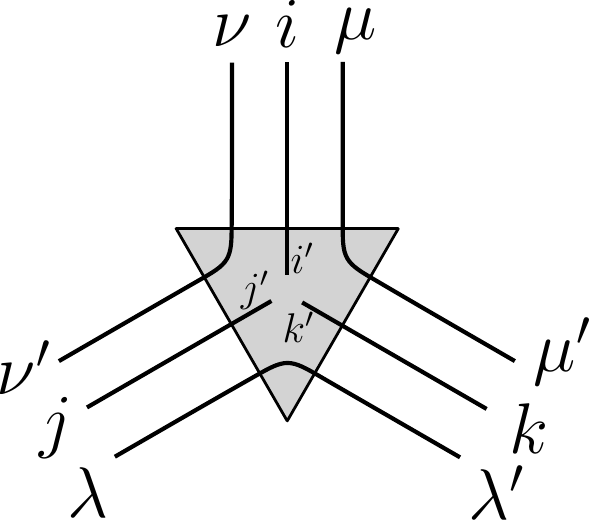}} =  \sqrt[6]{ \tilde{d}_\lambda \tilde{d}_\mu \tilde{d}_\nu} \, \sqrt[4]{d_i d_j d_k} \,G^{i j k}_{\lambda \mu \nu} \; \delta_{i i'} \delta_{j j'} \delta_{k k'} \delta_{\lambda \lambda'} \delta_{\mu \mu'} \delta_{\nu \nu'} , 
		\end{equation}
		where $ i' $, $ j' $ and $ k' $ represent the physical degrees of freedom.
		Each virtual index is associated to two physical indices (appearing in the tensors at the vertices connected by that edge). The first three $ \delta $-functions in Eq.~\eqref{eq:PEPS} guarantee that these two physical indices are always equal.
		The PEPS tensor for the vertices with the inverse orientation is obtained by rotating Eq.~\eqref{eq:PEPS}.

		\subsection{Ansatz for \texorpdfstring{$\theta \in [-\pi/2,0]$}{region 2}}
			For $\theta \in [-\pi/2,0]$, the ansatz is defined similarly:
			\begin{align}\label{eq:ansatzneg}
				\ket{\gamma}_{-} &= \mathcal{N} \prod_{p} (\id + \gamma d_1 O_1^p) \ket{1} ,\\
					&= \mathcal{N} \prod_{p} \left ( \sum_i  \tilde{d}_i O_i^p \right ) \ket{1}  .
			\end{align} 
			Note that we are now acting on the $ \ket{1} = \otimes_l \ket{1}_l$ product state, as opposed to the $ \ket{0} $ state that we used for $\theta \in [0,\pi/2]$.
			We can use the same graphical representation of this state:
			\begin{equation}\label{eq:plaquettes_loops_tau}
				\ket{\gamma}_{-} = \mathcal{N} \sum_{\mu_1 , \mu_2 , \dots} \tilde{d}_{\mu_1} \tilde{d}_{\mu_2} \dots \raisebox{-2cm}{\includegraphics[scale=.3]{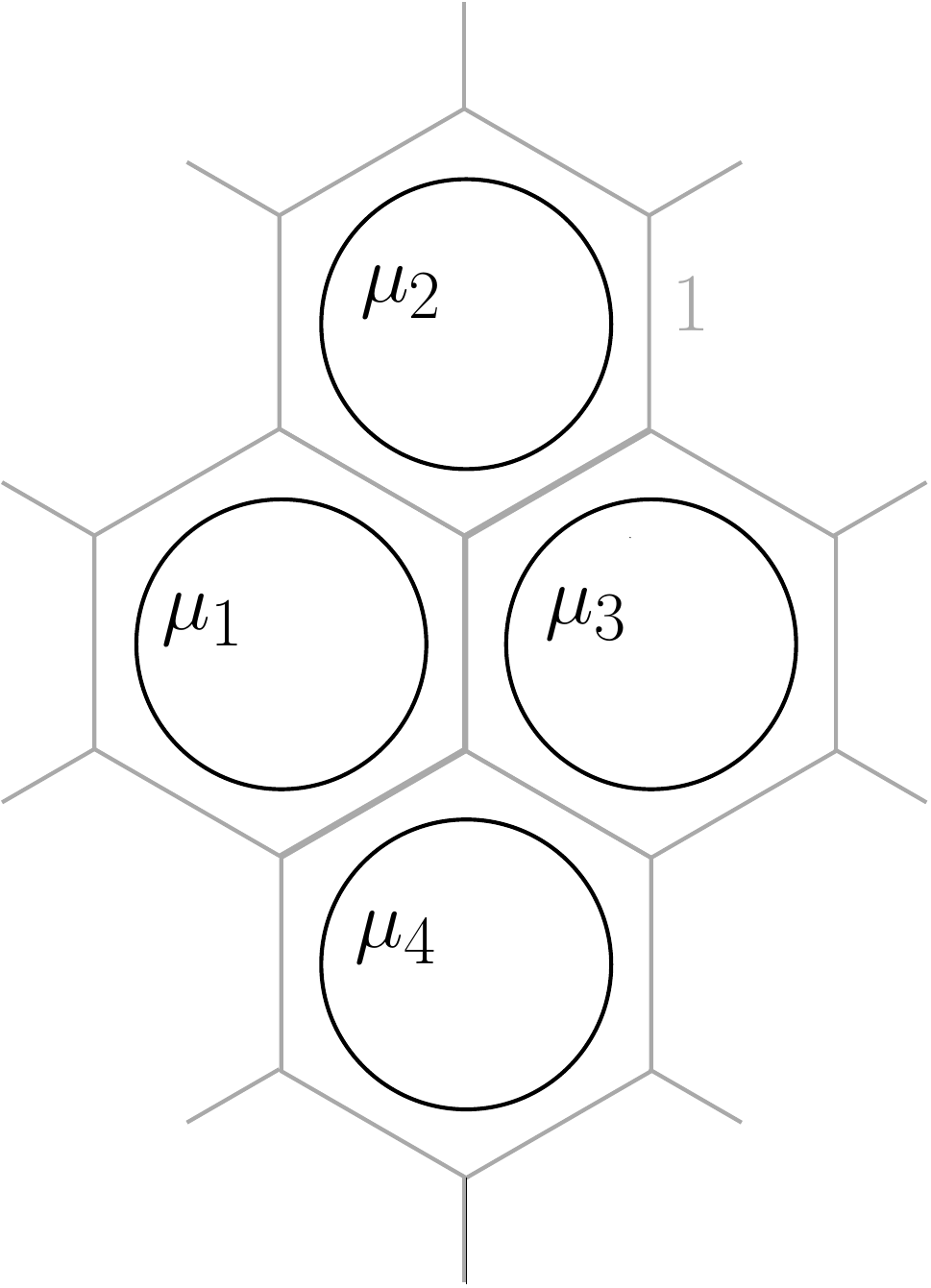}},
			\end{equation}			
			where the gray lines are now in the $ \ket{1} $ state initially.			
			As done for $\theta \in [0,\pi/2]$, we first fuse the loops along every edge:
			\begin{align}\label{eq:munutau}
				\raisebox{-0.4cm}{\includegraphics[scale=.5]{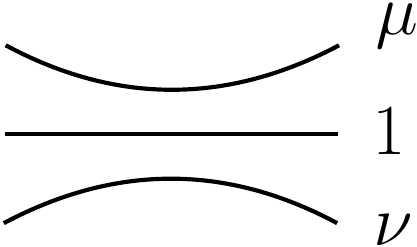}} & = \sum_{x} F^{1 1 \, 0}_{\mu\mu \, x} \;\; \raisebox{-0.4cm}{\includegraphics[scale=.5]{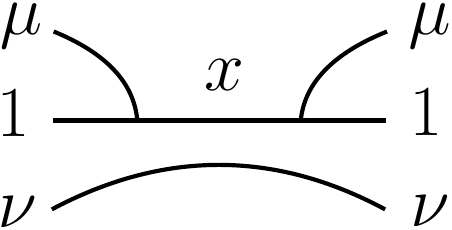}} ,\\
				& = \sum_{x} \sum_{k} F^{1 1 \, 0}_{\mu\mu \, x} F^{x x 0}_{\nu\nu k} \;\; \raisebox{-0.4cm}{\includegraphics[scale=.5]{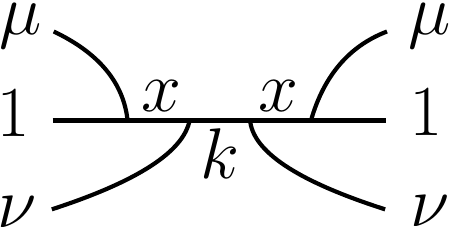}} \nonumber ,\\
				& = \sum_{x} \sum_{k} \sqrt{\dfrac{d_k}{d_1 d_\mu v_\nu}} \delta_{\mu 1 x} \delta_{x\nu k} \;\; \raisebox{-0.4cm}{\includegraphics[scale=.5]{munutau3.pdf}} \nonumber  .
			\end{align}
			The vertices then look like
			\begin{equation*}\label{key3}
					\raisebox{-1cm}{\includegraphics[scale=.45]{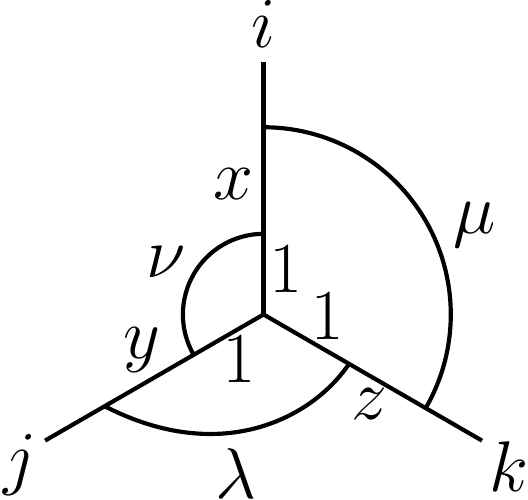}} \hspace{1cm} \text{and} \hspace{1cm} \raisebox{-1cm}{\includegraphics[scale=.45]{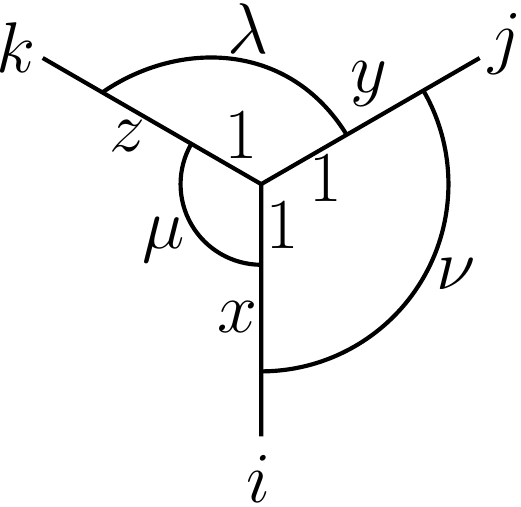}} .
			\end{equation*}
			To finish resolving everything into the lattice, these objects must be reduced to trivalent vertices. This is done by applying Eq.~\eqref{eq:vertex_loop} multiple times:
			\begin{equation}
				\raisebox{-.9cm}{\includegraphics[scale=.45]{vertextau_down.pdf}} = \sqrt{d_1^3 \, d_\lambda d_\mu d_\nu \,d_x d_y d_z }\, G^{x y 1}_{1 1 \nu} G^{x j z } _{\lambda 1 y} G^{i j k}_{z \mu x} 	\raisebox{-.6cm}{\includegraphics[scale=.4]{vertex.pdf}} ,
			\end{equation}
			and analogously:
			\begin{equation}
				\raisebox{-1.3cm}{\includegraphics[scale=.45]{vertextau_up.pdf}} = \sqrt{d_1^3 \, d_\lambda d_\mu d_\nu \,d_x d_y d_z } \, G^{z x 1}_{1 1 \mu} G^{k x y } _{1 \lambda z} G^{i j k}_{y x \nu} 	\raisebox{-.7cm}{\includegraphics[scale=.4]{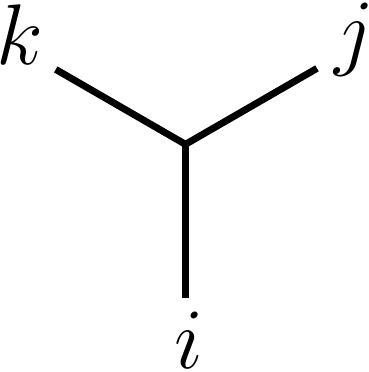}} .
			\end{equation}
			
			By splitting the factors appearing in Eq.~\eqref{eq:munutau} equally between the two adjacent vertices, we obtain the following PEPS tensors:
			\begin{equation}\label{eq:PEPS_tau_down}
				\raisebox{-1cm}{\includegraphics[scale=.42]{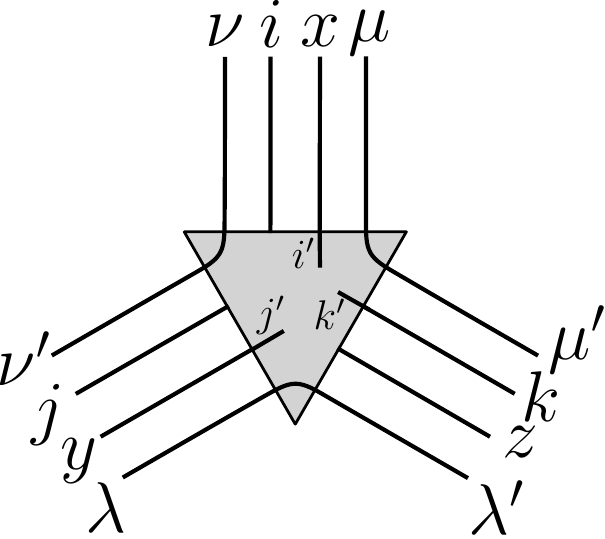}} =  \sqrt[6]{ \tilde{d}_\lambda \tilde{d}_\mu \tilde{d}_\nu}  \sqrt[4]{d_i d_j d_k}  \sqrt{d_x d_y d_z}\, d_1^{3/4} \, G^{x y 1}_{1 1 \nu} G^{x j z } _{\lambda 1 y} G^{i j k}_{z \mu x} \; \delta_{i i'} \delta_{j j'} \delta_{k k'} \delta_{\lambda \lambda'} \delta_{\mu \mu'} \delta_{\nu \nu'},
			\end{equation}
			
				\begin{equation}\label{eq:PEPS_tau_up}
				\raisebox{-1cm}{\includegraphics[scale=.42]{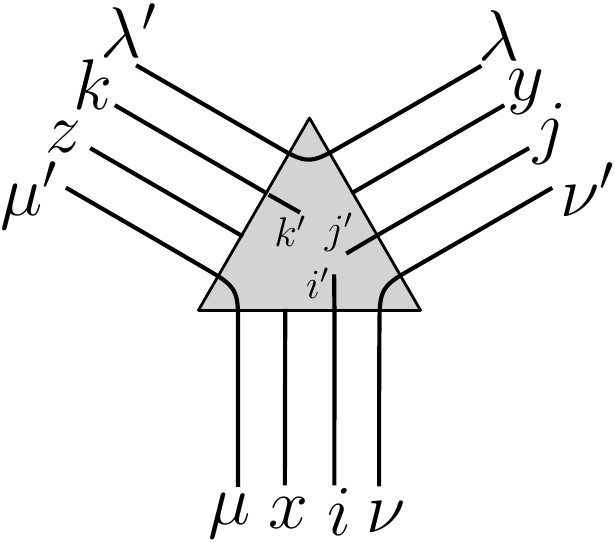}} =  \sqrt[6]{ \tilde{d}_\lambda \tilde{d}_\mu \tilde{d}_\nu} \sqrt[4]{d_i d_j d_k} \sqrt{d_x d_y d_z}\, d_1^{3/4} \, G^{z x 1}_{1 1 \mu} G^{k x y } _{1 \lambda z} G^{i j k}_{y x \nu}  \; \delta_{i i'} \delta_{j j'} \delta_{k k'} \delta_{\lambda \lambda'} \delta_{\mu \mu'} \delta_{\nu \nu'}.
				\end{equation}
				Note that there is one more virtual leg per side compared to Eq.~\eqref{eq:PEPS}.
				This is due to the extra sum appearing in Eq.~\eqref{eq:munutau}.
				
				Although, for $ \gamma=1 $,  the physical states given in Eqs.~\eqref{eq:ansatz} and \eqref{eq:ansatzneg} are identical, the PEPS tensors representing these two  states have very different properties. Indeed,  the double-layer transfer matrix with PEPS tensors \eqref{eq:PEPS_tau_down} and \eqref{eq:PEPS_tau_up} appears to be critical with a central charge which is twice that of the three-state Potts model while the PEPS tensor given by Eq.~\eqref{eq:PEPS} does not share this property. This observation clearly deserves further investigations.			

		\subsection{Reducing the bond dimensions}
			The PEPS tensor defined in Eq. \eqref{eq:PEPS} has a bond dimension (both physical and virtual) of $ 2^3 $.
			However, the $G$-symbol present in its definition, imposes certain rules which need to be met for the tensor to take a non-zero value. 
			These rules can be exploited to rewrite this tensor as one with a lower bond dimensions: for the $\mathbb{Z}_2$ theory \eqref{eq:Z2}, the bond dimension can be reduced to 4, while for the Fibonacci theory \eqref{eq:Fib} it can be reduced to 5. 
			
			The structure of the PEPS tensor, can also be exploited to reduce the bond dimension of the double-layer transfer matrix MPO tensors.
			Using the more efficient encoding of the tensor we just mentioned, the double layer bond dimension already gets reduced from 64 to 16 for $ \mathbb{Z}_2 $ and to 25 for Fibonacci.
			The Kronecker delta functions appearing in the right-hand side of Eq.~\eqref{eq:PEPS} allow us to further reduce these to 8 and 13 respectively.\\
			
			The same tricks can be applied to the tensors \eqref{eq:PEPS_tau_down} and \eqref{eq:PEPS_tau_up} (note that we only use this ansatz for the Fibonacci theory).
			The virtual bond dimension can be reduced from $ 2^4 $ to 8, while the physical bond dimension can be reduced from $ 2^3 $ to 5.
			The bond dimension of double-layer MPO tensors can be reduced from $ 2^8 $, to 34.\\
			
			As mentioned in the main body of this paper,  the variational energy per plaquette \eqref{eq:e0} is calculated using the VUMPS algorithm.
			Due to memory constrains, the bond dimension of the boundary MPS has to be limited to 100 for the $ \mathbb{Z}_2 $ model, and for the Fibonacci model with $\theta \in [0,\pi/2]$.
			Due to the higher bond dimension of the double-layer MPO obtained from ansatz \eqref{eq:PEPS_tau_down} and \eqref{eq:PEPS_tau_up}, the bond dimension of the boundary MPS has to be limited to 60 for the Fibonacci model with $\theta \in [-\pi/2,0]$.
			
\twocolumngrid

\end{document}